# Design and Fabrication of High Numerical Aperture and Low Aberration Bi-Convex Micro Lens Array

Jhy-Cherng Tsai[1], Ming-Fong Chen[1], and Hsiharng Yang[2]
[1]Deaprtment of Mechanical Engineering and [2]Institute of Precision Engineering
National Chung Hsing University
Taichung 40227, Taiwan

*Abstract-* **Micro lens array is crucial in various kinds of optical and electronic applications. A micro lens array with high numerical aperture (NA) and low aberration is in particular needed. This research is aimed to design and fabricate such a micro lens array with simple structure while keeps the same NA of a same-diameter hemisphere lens. A bi-convex semispherical micro lens array, with corresponding NA 0.379, by PDMS is first designed and analyzed. Experiments are further conducted to fabricate the designed micro lens array by the thermal reflow process. The formed profile is then sputtered with copper to serve as the mold. The front and the rear micro lens array are fabricated by plating PDMS to the mold and then assembled to form the designed micro lens array.**

## I. INTRODUCTION

Lens array is one of the most common devices employed in various kinds of optical and electronic applications, including light scattering or focusing and image processing as well as information storage. While dimensions of optical devices are highly reduced, micro lens array is in particular crucial for such applications. One way to reduce the dimension while keeping the focal length of a lens is to increase its numerical aperture (NA). Lens with high NA reduces focal spot in a short distance thus can effectively reduce the size of the data read/write mechanism for optical storage devices. NA, however, is related to the geometry, particularly the curvature, and the refractivity of the lens. Although hemispherical lens provides high NA, the image quality is relatively poor due to its high aberration that reduces the resolution of image. A micro lens array with high NA and low aberration is in particular needed as it features narrow focal spot, high lamination, and high resolution [1-5]. Many efforts have been employed, including non-spherical lens, Fresnel lens, and reflective-diffractive compound lens, to improve aberration. However, fabrication costs for these lenses are relatively high as they require special fabrication processes [6-8]. Among many different processes to fabricate such a high-NA low-aberration micro lens array, thermal reflow process preceded with lithography is the most economic approach [9, 10]. Thermal reflow process, on the other hand, is suitable for fabricating semispherical lens in small size due to the interfacial tension between the substrate and the material.

This research is aimed to design and fabricate a micro lens array with simple structure while keeps the same NA of a hemisphere lens of the same diameter. We start with the configuration design and analysis of such a high-NA and low-aberration microlens array with its corresponding fabrication processes and costs in mind. Experiments to fabricate a prototype of the designed microlens array is then designed and conducted as a proof of concept.

## II. NUMERICAL APERTURE AND ABERRATIONS

Fig. 1 shows the sketch and the geometry of a spherical lens. The numerical aperture of a spherical lens is related to its diameter $D$ and height $h$ and can be expressed as in equation (1) where $n_l$ is the reflectivity of the lens material.

$$NA = \frac{4Dh(n_l - 1)}{D^2 + 4h^2} \quad \quad (1)$$

The aberration of a lens includes spherical aberration, coma, astigmatism, curvature of field, and distortion. Fig. 2 illustrates the spherical aberration for parallel light beam. The light beam focus is based on the distance of traveling. As a result, the beam focus locates at different positions as shown is equation (2). The longitudinal and transverse aberrations, LA'$_R$ and TA'$_R$, are the difference between the real and expected locations as the shown in Fig. 2 [11].

$$\begin{aligned}
x' &= A_1 \sin\theta + B_1 s^3 \sin\theta + B_2 s^2 h \sin 2\theta + (B_3 + B_4) sh^2 \sin\theta + C_1 s^5 \sin\theta + C_3 s^4 h \sin 2\theta \\
&\quad + (C_5 + C_6 \cos^2\theta) s^3 h^2 \sin\theta + C_9 s^2 h^3 \sin 2\theta + C_{11} h^4 \sin\theta + \ldots \\
y' &= A_1 s \cos\theta + A_2 h + B_1 s^3 \cos\theta + B_2 s^2 h (2 + \cos 2\theta) + (3B_3 + B_4) sh^2 \cos\theta + B_5 h^3 \\
&\quad + C_1 s^5 \cos\theta + (C_2 + C_3 \cos 2\theta) s^4 h + (C_4 + C_6 \cos^2\theta) s^3 h^2 \cos\theta + (C_7 + C_8 \cos 2\theta) s^2 h^3 \\
&\quad + C_{10} s h^4 \cos\theta + C_{12} h^5 + \ldots
\end{aligned} \quad (2)$$

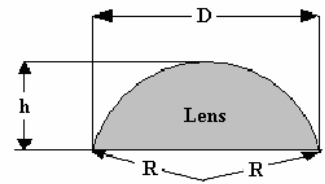

Fig. 1: Sketch of a spherical lens.

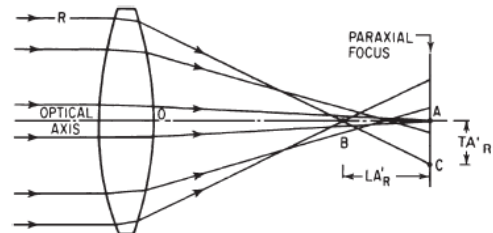

Fig. 2: Illustration of the spherical aberration. [11]

### III. CONFIGURATION DESIGN

As spherical lens provides high NA and easy to be fabricated with low cost, we take the advantage to increase the NA by increasing the curvature, *i.e.* reducing the diameter, of the lens. The aberration of such a lens also increases at the same time. We then put another spherical lens, convex to the other direction, to improve the aberration. The idea is shown in Fig. 3 where (a) illustrates the reduction of focal length and (b) presents the reduction of aberration with losing the focal length. The corresponding NA and spot size of such a bi-convex lens are shown in equations (3) and (4).

$$NA = \frac{D(n_l - 1)}{2} \left[ \frac{1}{R_1} - \frac{1}{R_2} + \frac{(n_1-1)t}{n_l R_1 R_2} \right] \quad \text{...............} (3)$$

$$Size = \frac{1.64\lambda}{D(n-1)\left[\frac{1}{R_1} - \frac{1}{R_2} + \frac{(n-1)t}{nR_1R_2}\right]} \quad \text{...............} (4)$$

Where D is the diameter; $R_1$ and $R_2$ are the radius of front and rear lenses; t is the thickness and n is the reflectivity of the lens.

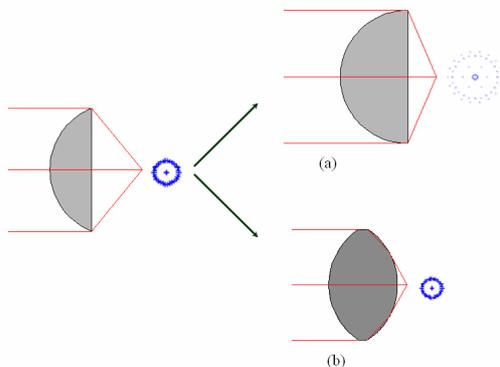

Fig. 3: Conceptual design of the bi-convex lens.

A dual-convex micro lens is designed with the target NA equals to a typical value for spherical lens of the same material. The diameter of the lens is expected to be less than 100μm. We then tune the radius of the front lens and the rear lens to improve the aberration for the He-Ne laser (λ=632.8nm). The best result shows the size and radius of the front and rear lenses are 97.6μm, 60μm, 79.7μm, and 43.5μm respectively, corresponding to sag-to-diameter ratios 0.17 and 0.2. Table I shows the comparison of aberration (ray fan, OPD fan and spot diagram) among the typical spherical lens, the non-spherical lens and the bi-convex lens for the same NA. Although the aberration is not as good as that of non-spherical lens, it is highly improved compared to the original spherical lens.

TABLE I
COMPARE OF ABERRATION FOR DIFFERENT TYPES OF LENS WITH NA=0.379

|  |  | Spherical lens | Non-spherical lens | Bi-convex lens |
|---|---|---|---|---|
| Aberration | ray fan | 500 | 0.01 | 5 |
|  | OPD fan | 50 | 0.01 | 1 |
|  | spot diagram (geo radius) | 299.25 | 0 | 2.671 |

### IV. PROTOTYPING AND RESULTS

Experiments are further conducted to fabricate the bi-convex micro lens array, designed in previous section, by the thermal reflow process as shown in Figure 4. The profile of the micro lens is dual semispherical thus can be easily fabricated. The photoresist AZ4620 is used in the lithography process. The cylindrical array after development is sent to the reflow process heated at 160°C for ten minutes. The formed profile is then sent to sputtering to serve as the mold. Micro lens array is then fabricated by plating PDMS to the mold. The front and the rear lens arrays are then assembled to form the designed bi-convex micro lens array.

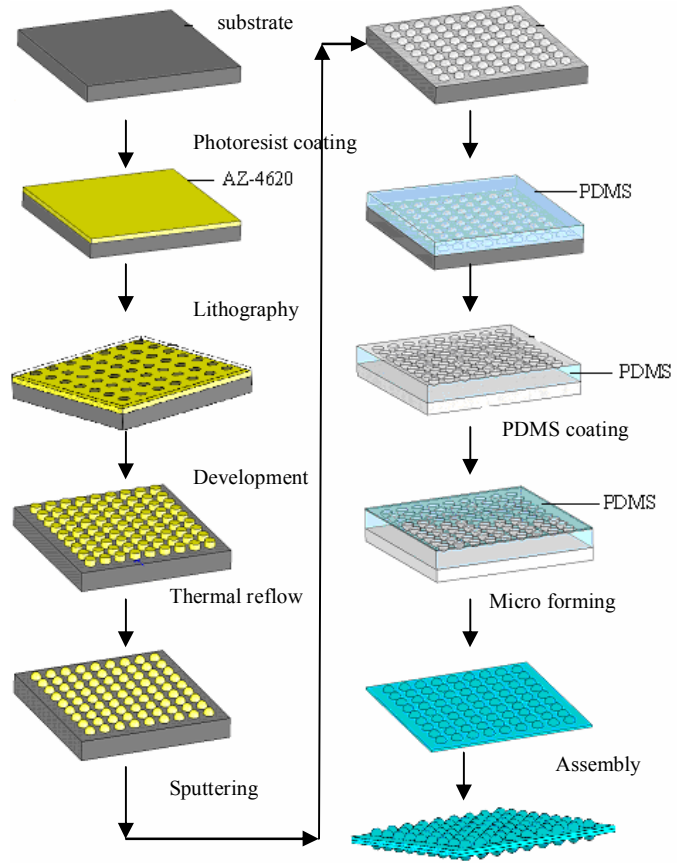

Fig. 4: fabrication processes of the designed bi-convex micro lens array.

The dimensions and surface profile of the photoresist after the thermal reflow process are measured by profiler. Average dimensions of the photoresist arrays are listed in Table II. The results show the repeatability of the fabrication process is good as the ratio of standard deviation to the average is acceptable for such a micro lens array. The average dimensions also show the accuracy is close to the designed target.

TABLE II
AVEREGAE DMENSIONS OF PHOTORESIST AFTER THERMAL REFLOW (UNIT: μm)

| Measured dimension | Front lens array | | | Rear lens array | | |
|---|---|---|---|---|---|---|
|  | $h_1$ | $D_1$ | $R_1$ | $h_2$ | $D_2$ | $R_2$ |
| average | 20.40 | 96.20 | 67.35 | 13.40 | 60.23 | 40.74 |
| Standard deviation | 1.60 | 0.17 | 4.26 | 0.75 | 0.37 | 1.63 |

The photoresist arrays are sent to sputtering to serve as molds for micro printing. Micro lens array are fabricated by platting PDMS on the mold. The fabricated front and rear micro lens arrays are measured by optical microscope, as shown in Fig. 5.

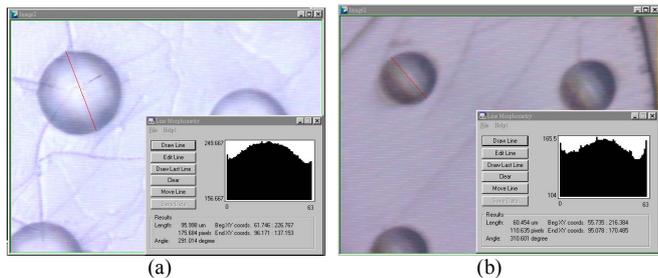

Fig. 5: Images of the fabricated (a) front and (b) rear micro lens array by optical microscope.

The diameters of the mold and the fabricated micro lens arrays are listed in Table III. The results show the repeatability of the sputtering and micro printing processes are very good as the ratio of standard deviation to the corresponding average of each mold and lens array is very tight for such a micro lens array. The average diameters showed the accuracy of each mold and fabricated micro lens array is fairly close to the designed target.

TABLE III
AVEREGAE DIAMETERS OF MOLD AND FABRICATED MICRO LENS ARRAY (UNIT: μm)

| Measured diameter | Mold | | Micro lens array | |
|---|---|---|---|---|
| | Front | Rear | Front | Rear |
| average | 95.7 | 63.0 | 95.6 | 61.3 |
| Standard deviation | 0.57 | 0.86 | 0.36 | 0.49 |

Finally, both the fabricated front and the rear micro lens arrays are aligned and assembled. The assembled bi-convex micro lens array is shown in Fig. 6(a). It is critical that both lens arrays have to be well aligned in order to produce the designed bi-convex micro lens array. A cross-hair mark was designed for this purpose. The mark was designed in both masks used in the lithography process. Both lens arrays are aligned when the cross-hair marks are superimposed as shown in Fig. 6(b).

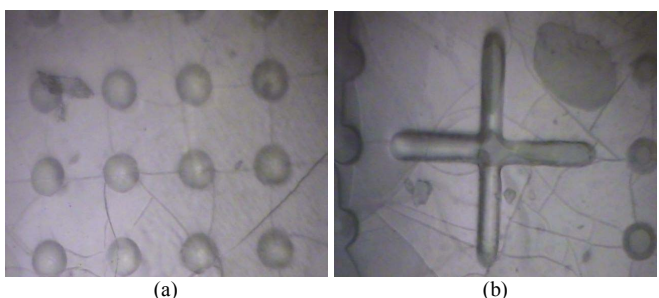

Fig. 6: (a) Image of the fabricated bi-convex micro lens array when (b) the cross-hair marks of both front and rear lens arrays superimposed.

## V. CONCLUSIONS

Micro lens array is a crucial component for use in various optical and electronic applications. This paper investigated the design and fabrication of a bi-convex micro lens array with high numerical aperture and low aberration. A bi-convex semispherical micro lens array, with corresponding NA 0.379, by PDMS is first designed and analyzed. Experiments are further planned and conducted to fabricate the prototype of the designed micro lens array by a process combining the lithography, thermal reflow, micro printing, and assembly processes. Experiment data show both the repeatability and the accuracy are acceptable good for such micro-scale lens array. Although the performance of the assembled micro lens array needs further inspection, the innovation and contribution in this research is to design and fabricate a prototype of a bi-convex micro lens array with high NA and low aberration.

ACKNOWLEDGMENT

This research has been supported by the National Science Council of Taiwan under contract NSC95-2221-E-005-014-MY3. The authors also want express their gratefulness to Professor Hsi-Fu Shih for his helpful discussions and Professor Gou-Jen Wang for his assistance on prototyping and measurement.